\documentclass{ws-ijmpd}

\usepackage{amssymb}
\usepackage{graphicx}
\usepackage{amsmath}
\usepackage{amsfonts}
\newcommand*{\be}{\begin{equation}}
\newcommand*{\ee}{\end{equation}}
\newcommand{\ba}{\begin{eqnarray}}
\newcommand{\ea}{\end{eqnarray}}
\newcommand{\bse}{\begin{subequations}}
\newcommand{\ese}{\end{subequations}}
\newcommand{\M}{{\cal {M}}}
\newcommand{\B}{{\cal {B}}}

\newcommand{\HH}{{\cal {H}}}

\begin{document}


%
\catchline{}{}{}{}{}
%

\title{\textbf{DYNAMICS OF A SELF-GRAVITATING MAGNETIZED SOURCE.}}

\author{A. Ulacia Rey\footnote{Present address: ICIMAF, Calle E No-309 Vedado, cp-10400,
Ciudad Habana. Cuba}}

\address{Facultad de Tecnolog\'{\i}a de la Salud, Dr. Salvador Allende, Cerro\\
Ciudad de la Habana, cp-10400, Cuba\\
alain@icmf.inf.cu}

\author{A. P\'{e}rez Mart\'{\i}nez}

\address{Instituto de Cibern\'{e}tica Matem\'{a}tica y F\'{i}sica (ICIMAF),Calle E No-309
Vedado\\ Ciudad de la Habana, cp-10400, Cuba\\
aurora@icmf.inf.cu}

\author{Roberto. A. Sussman}

\address{Instituto de Ciencias Nucleares, Universidad Nacional Aut\'{o}noma de M\'{e}xico
(UNAM)\\ Distrito Federal 04510, cp-70543, M\'{e}xico\\
sussman@nuclecu.unam.mx}

\maketitle

\begin{history}
\received{Day Month Year}
\revised{Day Month Year}
\comby{Managing Editor}
\end{history}

\begin{abstract}
We consider a magnetized degenerate gas of fermions as the matter
source of a homogeneous but anisotropic Bianchi I spacetime with a
Kasner--like metric. We examine the dynamics of this system by
means of a qualitative and numerical study of Einstein-Maxwell field
equations which reduce to a non--linear autonomous system. For all
initial conditions and combinations of free parameters the gas
evolves from an initial anisotropic singularity into an asymptotic
state that is completely determined by a stable attractor.
Depending on the initial conditions the anisotropic singularity is
of the ``cigar'' or ``plate'' types.
\end{abstract}


\section{Introduction}
Intense  magnetic fields may play an important role in astrophysical
objects, such as compact object like neutron or white dwarf stars.
Weaker magnetic fields~\cite{A} may also be important in connection
with the magnetic plasma effects associated with local anisotropy in
self-gravitating and collision-less systems (e.g., galactic halos of
fermionic dark mater~\cite{bf}) and net radiation flow entering (or
leaving) the gas clouds. Strong magnetic fields can also arise as
plausible explanation behind the origin of anisotropy in highly
dense systems, like solid core, exotic phase transitions, pion
condensation or anisotropic compact object by ultra-strong
self-magnetization $(\B>10^{14})$ Gauss. The latter phenomenon
motivates the present work, as a theoretical explanation behind the
anisotropy of stresses in these systems. This is connected with
previous work~\cite{V,Lattimer,A,M} in which we studied the
thermodynamical properties of gases of electrons and vectorial
bosons using the Electroweak model. We showed in that work that such
a system was characterized by an anisotropic energy--momentum tensor
in which the pressure transverse to the magnetic field ($p_{\perp}$)
was (in general) different from that parallel to the field
($p_{\parallel}\ne p_{\perp} $). Under a Newtonian framework, we
found that the vanishing of $p_{\perp}$ would drive the system to
gravitational collapse. We also discussed if critical values of the
magnetic field exist, allowing for the vanishing of this pressure
and its subsequent collapse. By assuming that the model could
describe a Newtonian compact object, we concluded that the final
state of such a star would  never be a magnetar of the type of
Duncan and Tompson~\cite{Tompson}. Instead, such a state could be a
strange star or a ``black cigar--like'' structure.

However, the proper study of highly magnetized sources
of possible compact objects requires that we incorporate
the non--linear effects of Einstein's General Relativity theory,
allowing us to examine the effects of the gas magnetization in the
collapse of a magnetized self-gravitating Fermi gas. Ideally, we
should consider minimally realistic rotating
configurations, but this would require elaborate
hydrodynamical numerical techniques. Instead and as a first step, we
consider simple and mathematically more tractable, even if extremely
idealized, configurations that could reveal (at least) some of the
properties of the dynamical evolution of magnetized gases in General
Relativity. We undertake such a study in the present article by
considering what could be the simplest non--static geometry compatible
with the anisotropies of a magnetized gas: the homogeneous but
anisotropic Bianchi I spacetime in a Kasner--like coordinate basis.

\section{Magnetized Fermi gas for a Kasner metric.}

The Bianchi I spacetime can be described by the Kasner--like metric
element~\cite{MTW,WE}:
\begin{equation}
{ds^2} \ = \
-c^2\,dt^2+A^2(t)\,dx^{2}+B^2(t)\,dy^2+C^2(t)\,d{z}^2,\label{Kasner}
\end{equation}
corresponding to an homogeneous, but anisotropic, spacetime with zero
spacial curvature and having all relevant quantities depending
only on time. We consider as a source of (\ref{Kasner})
a degenerated magnetized Fermi gas whose energy--momentum tensor is
that obtained by {A. P\'{e}rez Mart\'{\i}nez and H.P\'{e}rez Rojas
\cite{Au}}.

The energy--momentum tensor for this gas in a comoving frame with
coordinates $x^0=ct,\,x,\,y,\,z$ and 4--velocity $u^a = \delta^a_0$
is given by
\begin{equation}
T^{a}_{b}=(U+P)\,u^{a}u_{b}+P\delta^{a}_{b}+\Pi^{a}_{b},  \qquad
P=p-\frac{2}{3}B\M,\label{Tab2}
\end{equation}
where $U$ is the total matter--energy density, $\B$ the magnetic field
and $\M$ the magnetization of the gas, $P$ is the isotropic
pressure and $\Pi^{a}_{b}$ is the traceless anisotropic pressure
tensor, which for (\ref{Kasner}) takes the form
$\Pi^{a}_{b}=\textrm{diag}[\Pi,\Pi,-2\Pi,0]$ with
$\Pi=-\frac{1}{3}\B\M$. Thus, $p=T^z_z$ is the stress tensor
component (pressure) in the direction of the magnetic field.

The full equation of state for this gas takes the following
form~\cite{V,M}:
\be p= \ \lambda\,\,\beta\, \Gamma_{p}(\beta,\mu), \ \ \ \B\M \ = \
\lambda\, \beta\, \Gamma_{{\M}}(\beta,\mu), \ \ \ U \ = \ \lambda\,
\beta\, \Gamma_{_U}(\beta,\mu). \ee
where the $\Gamma$ functions are given further ahead, $\beta$, the
magnetic field normalized with a suitable critical value $\B_c$,
and $\lambda$ are given by:
\be \beta \ = \ \frac{\B}{\B_{c}},\label{Gamma} \ \ \ \lambda \ =
\ \frac{mc^2}{4\pi^2\lambda_{c}^3}.\label{lambda} \ee
If we consider our gas to be made of by electrons, then $m$ is the
electron mass, $m_e$, while $\lambda_{c}=\hbar/mc$ is the Compton
wavelength and $B_{c}=m^2c^3/e\hbar$ is the so--called critical
magnetic field \footnote{(e.g. for the electron
$\B^{e}_{c}=4,41\times 10^{13}$ Gauss)}. In this limit of strong
magnetic field, the functions $\Gamma(\beta,\mu)$ take the form:
\be\label{Gs} \Gamma_{p}= \sum\limits_{n = 0}^s \alpha_{n}
(a_{n}-b_{n}-c_{n}), \ \
 \Gamma_{_{\M}}=\Gamma_{p}
-\sum\limits_{n = 0}^s\alpha_{n}c_{n}, \ \
 \Gamma_{_U}=-\Gamma_{p}+
2\sum\limits_{n = 0}^s\alpha_{n}a_{n} \ee
where $a_{n}={\mu\sqrt{\mu^2-1-2n\beta}}$,
$b_{n}=\ln[(\mu+a_{n}/\mu)/\sqrt{1+2n\beta}]$, $c_{n}=2n\beta
b_{n},$ $\alpha_{n}=2-\delta_{0n}$, and $\mu=\mu_{e}/(m_{e}c^2)$
chemical potential of the fermions,
$s=I[\frac{1}{2}(\B_c/\B)(\mu^2-1)]$ corresponds to the maximum
Landau level for a given Fermi energy and magnetic field strength,
while $I[X]$ denotes the integer part of its argument $X$.

Although (\ref{Kasner}) is obviously inappropriate to
examine the magnetized gas as a source of a compact object, we will
consider in our analysis the conditions prevailing inside a very high
density compact object in which we can ignore the gas temperature.

\section{ Einstein--Maxwell field equations.}

The comoving 4-velocity $u^{a}=\delta^a_0$ in (\ref{Kasner}) yields
the expansion scalar $\Theta =u^a\,_{;a}$ and the traceless shear
tensor $\sigma_{ab}=u_{(a;b)}-(1/3)\Theta h_{ab}$. In order to
obtain a self-consistent system of ordinary differential equations
suitable for numerical integration, it is convenient to eliminate
second order derivatives of the metric functions by expressing the
coupled Einstein-Maxwell field equations $G^{a}_{b}=\kappa
T^{a}_{b}$, $F^{ab}\,_{;b}=0$ and $F_{[ab;c]}=0$, as well as  the
balance equations $T^{ab}\,_{;b}=0$, in terms of $\Theta$ and
$\sigma^a_b$. Following this approach and after tedious algebraic
manipulations, we can reduce the field and balance equations to the
following self--consistent system of 4 independent ordinary
differential equations:
\bse\label{efe2}\ba \dot U \ &=& \
-(U+p-\frac{2}{3}\B\M)\,\Theta-\B\M\Sigma
\\
\dot \Sigma \ &=& \ \frac{2}{3}\kappa\, \B\M-\Theta\,\Sigma
\\
\dot\Theta \ &=& \ \kappa \,(\B\M+\frac{3}{2}(U-p))-\Theta^2
\\
\dot\beta \ &=& \ \frac{2}{3}\beta (3\Sigma-2\Theta) \ea\ese
where $\Sigma\equiv \sigma^z_z$ and a dot denotes derivative with
respect to $ct$. From (\ref{Gs}), the state variables $U,\,p$ and
$\B\M$ depend on the normalized magnetic field and chemical
potential, $\beta$ and $\mu$.

It is convenient to assume the zero order approximation ($n=0$), so
that we would be considering all electrons in the basic Landau
level~\cite{V}. Under this approximation the expansions of the
$\Gamma(\beta,\mu)\equiv \Gamma(\mu)$ functions in (\ref{Gs})
simplify considerably: $\Gamma_p=\Gamma_{_\M} =a_0-b_0$\, and \,
$\Gamma_{_U}=a_0+b_0$,\, where $a_0,\,b_0$ are given by setting
$n=0$ in (\ref{Gs}).

Introducing the following new variables and dimensionless functions:
\ba H \ = \ \frac{\Theta}{3},\qquad
\frac{d}{d\,\tau} \ = \ \frac{1}{H}\,\frac{d}{c\,dt},\qquad
S=\frac{\Sigma}{H}, \qquad \Omega=\frac{\kappa\lambda\beta}{3H^2},
\qquad  \HH \ = \ \frac{H}{H_0},\label{defs2}
\ea
where $H_0$ is a constant with inverse length units.
Einstein--Maxwell equations (\ref{efe2}) become the following
dimensionless system
\bse\label{efe4}\ba \Omega' \ &=& \ 2
\left\{1+S-\left[\mu\sqrt{\mu^2-1}+2\ln\left(\mu+\sqrt{\mu^2-1}\right)
\right]\Omega
\right\}\,\Omega,\label{efe4a}\\
S' \ &=& \
\left[(2-S)\,\mu\sqrt{\mu^2-1}-2(1+S)\,\ln\left(\mu+\sqrt{\mu^2-1}
\right)\right]\Omega,\label{efe4b}\\
\mu\,' \ &=& \
\frac{\left[(2-S)\,\ln\left(\mu+\sqrt{\mu^2-1}\right)-3S\,\mu\sqrt{\mu^2-1}
\right]}{2\mu^2}{\sqrt{\mu^2-1}},\label{efe4c} \ea\ese
supplemented by
\begin{equation}\HH' \ = \
\left\{-3+\left[\mu\sqrt{\mu^2-1}+2\ln\left(\mu+\sqrt{\mu^2-1}\right)\right]
\,\Omega\right\} \HH.\label{eqH}\end{equation}
where a prime denotes derivative with respect to the dimensionless
time $\tau$ and we have chose a length scale characterized by
setting $({3\,H_0^2}/{\kappa\,\lambda})  \ = \  1$, so that (from
(\ref{defs2})), we obtain
\begin{equation}\beta \ = \ \Omega\,\HH^2.\label{betaOm}\end{equation}
Since we have $\kappa\lambda=0.749\times 10^{-24}\textrm{cm}^{-2}$
for electrons, this choice leads to $1/H_{0}=2\times
10^{12}\,\textrm{cm}$, which is much smaller than the cosmological
Hubble radius. \footnote{It is of the order of the distance from the
Earth to the Sun.}

\section{Qualitative and numerical analysis.}

The system (\ref{efe4}) defines a three dimensional phase space
$(\Omega,S,\mu)$ with $\Omega\geq 0$ and $\mu\geq 1$. The planes
$\Omega= 0$ and $\mu=1$ are invariant subsets corresponding to a
Kasner vaccum. Besides these subsets, we also have:
\bse\ba \textrm{I}&=&\left\{\mu=1,\, S=-1,\, \Omega\right\}\\
\textrm{II} &=& \left\{\mu,\, S(\mu)={\frac {{2\,b_0}\, \left(
-1+2\,{\mu}^{2}+2\,{a_0} \right) }{
 \left( -3+6\,{\mu}^{2}+2\,{ b_0} \right) {a_0}+ \left( -1+2\,{\mu}^{
2} \right) {
b_0}-6\,{\mu}^{2}(1-\,{\mu}^{2})}},\,\Omega\right\},\nonumber\\
\\\textrm{III} &=& \left\{\mu=1.42,\, S=0.34,\,\Omega=0.42\right\}
\ea\ese
where $a_0,\,b_0$ are given by setting $n=0$ in (\ref{Gs}). The
curves I\, and\, II are saddles, while III is a stable attractor. As
shown in figure 1, some trajectories start in a singular state
characterized by $\Omega\to\infty$ while others start in a singular
Kasner vacuum $\Omega=0$, though all terminate in the attractor III.
Depending on initial conditions, some trajectories approach eras of
Kasner vacum I,\,II. The asymptotic state is given by III, which
(from (\ref{betaOm})) corresponds to $\beta\to 0.42\,\HH^2$.
Numerical examination of (\ref{efe4}) and (\ref{eqH}) reveals that
$\beta$ has a an asymptotic power law decay in terms of the physical
time $t$. Trajectories that initiate in a singular Kasner vacum
denote an unphysical evolution passing from $\Omega=0$ to the value
$\Omega=0.42$ given by the attractor III. The most physically
motivated trajectories correspond to those in which $\Omega$
decreases monotonically to the asymptotic value in III, hence the
normalized magnetic field $\beta$ also exhibits monotonous decay
approximately following a power law scaling law (in terms of the
physical time $t$).

\begin{figure}
\centering
\includegraphics[height=10cm,width=8cm,angle=270 ]{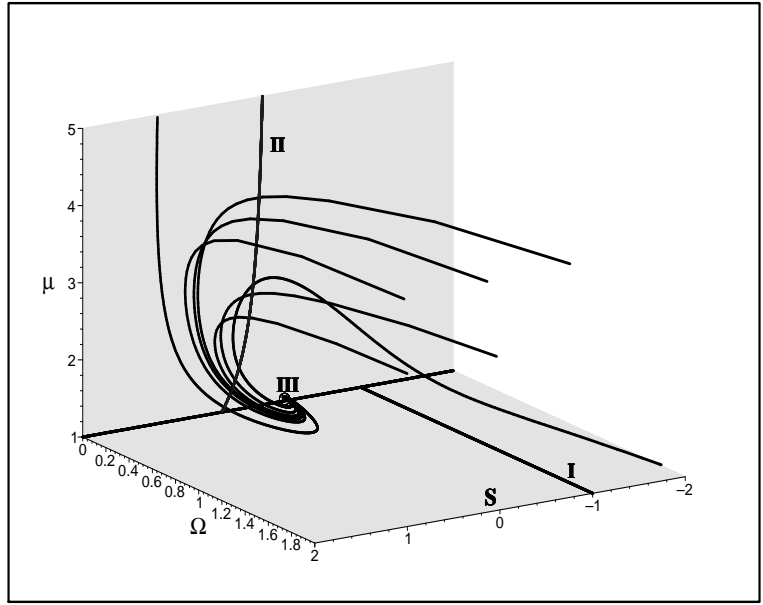}
\caption{ Phase space $(\Omega,S,\mu)$. Notice how all trajectories
evolve towards the attractor III. Some curves approach states of
Kasner vacum (the saddles $\Omega=0$, I and II). Some curves have an
initial singular vacum state (III) and some initiate at
$\Omega\to\infty$. Singular state are characterized by anisotropic
singularities. }
\label{fig1}       
\end{figure}

Since the physically interesting range of magnetic field intensities
in a magnetized electron gas is $\B\geq \B_c $  (or $\beta\geq 1$),
it is important to examine the behavior of the trajectories near the
initial singular states. The nature of the initial singularity can
be appreciated by looking at the behavior of the scale factors
$A,\,B,\,C$ in (\ref{Kasner}) in the limit $\tau\to-\infty$. The
initial singularity (whether a Kasner vaccum or not) is generically
anisotropic and can be of the ``cigar'' or ``plate'' types. In the
former case we have $C\to\infty$ with diverging $\Omega$, while both
$A,\,B$ tend to zero, forming a singular line along the $z$
direction parallel to the magnetic field.

\section{Conclusion}

We have examined the evolution of magnetized degenerate gas of
fermions, considering the case in which the fermions are electrons
in their basic Landau level ($n=0$ in (\ref{Gs})). Einstein--Maxwell
equations for a Bianchi I spacetime yield the dynamical system
(\ref{efe4}). Initial conditions exist for an evolution starting in
a line singularity with $\Omega$ (and $\beta$) decaying monotonously
to the asymptotic value in the attractor III. This type of evolution
is similar to that emerging in previous Newtonian
studies~\cite{M,A}. However, we have also the possibility of a
non--vaccum initial singular state that is not unphysical (diverging
$\Omega$) and is characterized by a ``plate'' type of singularity,
which is a counter--intutive type of anisotropy that lacks a
Newtonian equivalent. A more detailed and extensive study of the
singular states in this gas, as well as a less restrictive form of
the equation of state and the case of a gas of neutrons (instead of
electrons), is currently under investigation.

\section{Acknowledgments}

We wish to acknowledge support from grant PAPIIT-DGAPA number
IN-117803, from the Caribbean Network for ICAC-ICTP.

\end{document}